\documentclass[journal=jacsat,manuscript=article]{achemso}

\usepackage{chemformula} 
\usepackage[T1]{fontenc} 
\usepackage{multirow}
\usepackage{titlesec} 
\titleformat{\section}{\normalfont\Large\bfseries}{\Roman{section}.}{1em}{} 

\author{Elham Ghobadpour}
\affiliation[Grenoble]
{CNRS, UMR 5525, VetAgro Sup, Grenoble INP, TIMC, Université Grenoble Alpes, Grenoble, France}
\alsoaffiliation[ENS]
{CNRS, ENS de Lyon, Laboratoire de Physique (LPENSL UMR 5672), 69342 Lyon Cedex 07, France}
\author{Max Kolb}
\affiliation[CBP]
{ENS de Lyon, Centre Blaise Pascal (CBPsmn), 69342 Lyon Cedex 07, France}
\author{Ivan Junier}
\affiliation[Grenoble]
{CNRS, UMR 5525, VetAgro Sup, Grenoble INP, TIMC, Université Grenoble Alpes, Grenoble, France}
\author{Ralf Everaers}
\email{ralf.everaers@ens-lyon.fr}
\affiliation[CBP]
{ENS de Lyon, CNRS, Laboratoire de Physique (LPENSL UMR 5672), 69342 Lyon Cedex 07, France}
\alsoaffiliation[CBP]
{ENS de Lyon, Centre Blaise Pascal (CBPsmn), 69342 Lyon Cedex 07, France}

\title[emergent_dynamics]
  {The emergent dynamics of double-folded randomly branching ring polymers}

\begin{document}

\begin{abstract}
The statistics of randomly branching double-folded ring polymers are relevant to the secondary structure of RNA, the large-scale branching of plectonemic DNA (and thus bacterial chromosomes), the conformations of single-ring polymers migrating through an array of obstacles, as well as to the conformational statistics of eukaryotic chromosomes and melts of crumpled, non-concatenated ring polymers. Double-folded rings fall into different dynamical universality classes depending on whether the random tree-like graphs underlying the double-folding are quenched or annealed, and whether the trees can undergo unhindered Brownian motion in their spatial embedding.
Locally, one can distinguish (i) repton-like mass transport around a fixed tree, (ii) the spontaneous creation and deletion of side branches, and (iii) displacements of tree nodes, where complementary ring segments diffuse together in space. Here we employ dynamic Monte Carlo simulations of a suitable elastic lattice polymer model of double-folded, randomly branching ring polymers to explore the mesoscopic dynamics that emerge from different combinations of the above local modes in three different systems: ideal non-interacting rings, self-avoiding rings, and rings in the melt state. We observe the expected scaling regimes for ring slithering, the dynamics of double-folded rings in an array of obstacles, and Rouse-like tree dynamics as limiting cases. Notably, the monomer mean-square displacements of $g_1\sim t^{0.4}$ observed for crumpled rings with $\nu=1/3$ are similar to the subdiffusive regime observed in bacterial chromosomes.
In our analysis, we focus on the question to which extent contributions of different local dynamical modes to the emergent dynamics are simply additive.
In particular, we reveal a non-trivial acceleration of the dynamics of interacting rings, when all three types of local motion are present. 
In the melt case, the asymptotic ring center-of-mass diffusion is dominated by the contribution from coupling of the ring-in-an-array-of-obstacles dynamics with the tree dynamics. This contribution scales inversely with the ring weight and is compatible with a scenario in which constraint release restores a Rouse-like dynamics. 
\end{abstract}

%
%
\section{INTRODUCTION}\label{sec:int}

\begin{figure}[t!]
\includegraphics[width=\columnwidth]{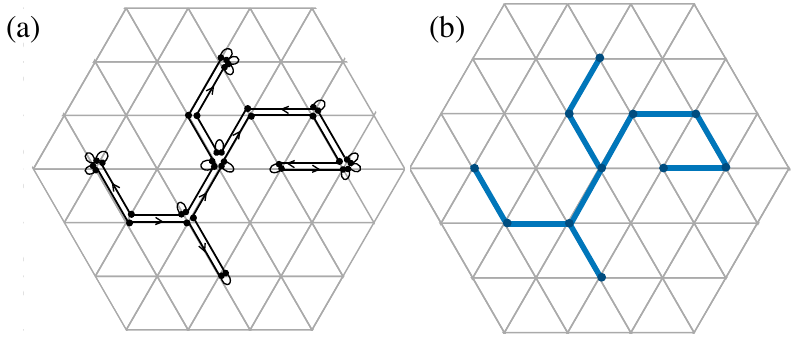}
\caption{(a) An example of a (tightly wrapped) double-folded ring polymer and (b) the corresponding tree on a trigonal lattice. Small loops represent reptons (bonds of zero length, see~Section~\ref{sec:Model and Method}), for which adjacent monomers along the ring occupy identical lattice sites.}
\label{fig:WrappedRing}
\end{figure}

Large ring polymers exhibit surprisingly rich behavior. 
When diffusing through an array of obstacles, they naturally adopt double-folded conformations, forming randomly branching polymer structures or trees~\cite{Khokhlov1985Phy.L.A, Rubinstein1986PhysRevLett., Obukhov.Rubinstein1994PRL,Grosberg2014SoftMatter,Rosa2014Phys.Rev.Lett.,SmerkGrosberg2015Cond.Matt} (Fig.~\ref{fig:WrappedRing}). 
Interestingly, double-folded tree-like ring conformations have broad significance in both physical~\cite{Kapnistos:2008NatMater} and biological systems. Biological examples include plectonemes in supercoiled DNA molecules~\cite{vinograd_twisted_1965,Vologodskii_review1994,MarkoSiggia1995} and, consequently, their role in bacterial chromosome structuring~\cite{DELIUS1974, Kavenoff:1976, Vologodskii_review1994,Junier:2023FrontMicrobiol}, large RNA molecules~\cite{RNALiu2016,vRNAKellyGrosberg2016,PrueferRNASingaram2016}, and
the arrangement of DNA in eukaryotic cells during interphase,~\cite{Grosberg_1993,Rosa2008PLOS,LiebermanAiden2009.Science,Halverson:Rep.Prog.Phys2014} which bears striking similarities to the behavior of melts of non-concatenated ring polymers.~\cite{Muller:1996PRE,BrownJCP1998,Halverson:2011JCP,Rosa2014Phys.Rev.Lett.,Schram2019SoftMatter}
Areas of current research~\cite{Kruteva:2023Macromolecules} include the full characterization of threading phenomena,~\cite{SmerkGrosberg2016ACSMacroLet.,Schram2019SoftMatter} where inter-ring penetrations impose long-lived topological constraints~\cite{Michieletto:2016ProcNatl} that dramatically slow down relaxation~\cite{Michieletto:2017Polymers,Smrek:2019ACSMacroLett, Ubertini:2022Macromolecules}, 
as well as the alternative scenario of activated dynamics induced by inter-ring caging~\cite{Schweizer:2023ACSPolym,Mei_PNAS:2024}.
Yet, the scaling of the center-of-mass diffusion coefficient \( D \) with ring length \( N \) in melts remains unresolved. None of the current theoretical predictions for the scaling exponent are consistent with molecular dynamics simulations, which observe slower diffusion~\cite{Halverson:2011JCP}.


Here we explore the dynamics of tightly double-folded using a model that does not allow for threading. Double-folded rings fall into different dynamical universality classes depending on whether the random tree-like graphs underlying the double-folding are quenched or annealed~\cite{Grosberg93Universality}, and whether the trees can undergo unhindered Brownian motion in their spatial embedding. Specifically, one can identify three distinct types of local dynamics: 

\begin{enumerate}
\item \textbf{Reptons} transporting ring mass along a fixed tree~\cite{RubinsteinRepton1987,MarkoSiggia1995} (Fig.~\ref{fig:MC_moves}(a)).  
\item \textbf{Creation or retraction of side-branches} where either (i) one of the two double-folded ring segments double-folds onto itself and branches away from the current tree or where vpnversly (ii) two antiparallel ring segments wrapping a side branch are retracted into the rest of the tree in the inverse process (Fig.~\ref{fig:MC_moves}(b)).
\item \textbf{Brownian motion of tree nodes} maintaining the tree connectivity and the secondary structure of the double-folded rings~\cite{Rosa2019The.EPJE} (Fig.~\ref{fig:MC_moves}(c)-(d)). 
\end {enumerate}

The effect of repeated local modifications (or ``moves'' in the Monte Carlo (MC) language we will adopt throughout the article) is not simply additive to the resulting dynamics of the ring and tree conformations. Instead, an \textit{emergent dynamics} typically arises because the ring ``monomers'' and tree ``nodes'' are not independent degrees of freedom but are coupled through (i) the ring connectivity, (ii) the constraint of double-folding restricting ring conformations to those with underlying tree structures, (iii) additional Hamiltonian terms governing interactions between monomers or nodes, branching activity, or chain stiffness, and (iv) local topological constraints arising from the mutual impenetrability of ring and tree backbones. The emerging internal dynamics
typically displays intermediate scaling regimes, whose understanding is important to
comprehend the long-time diffusion constant.

This coupling is not a particularity of rings. In linear chains, the connectivity between successive monomers gives rise to the characteristic Rouse intrachain dynamics, while topological constraints are at the origin of the reptation dynamics in the tube model~\cite{DoiEdwards}. Dynamical analogues can be formulated for rings. In particular, a well-understood limit is the ``ring-in-an-array-of-obstacles'' (RiAO) dynamics~\cite{Rubinstein1986PhysRevLett.,SmerkGrosberg2015Cond.Matt,Rubinstein:Macromolecules2016}, which arises from the coupling of local mass transport along the tree and the creation and deletion of side branches. However, understanding the general case remains a challenge, more particularly in the melt case, where the ``obstacles'' represent co-localized ring segments that are not fixed in space but instead have a dynamics of their own~\cite{Rubinstein:Macromolecules2016}. 


The purpose of the present paper is to pave the way for solving this challenge by introducing: 
\begin{itemize}
\item a model system, where the various interactions and types of local dynamics can be switched on and off in a controlled manner,
\item a procedure to analyze the independency of contributions from different dynamic modes to the experimentally observable monomer and the center-of-mass mean-square displacements. 
\end{itemize}

The manuscript is structured as follows: 
In Section~\ref{sec:theory} we introduce the relevant observables and associated exponents, along with a brief summary of the theoretical background. 
In Section~\ref{sec:Model and Method}, we introduce our model and the corresponding dynamic MC scheme to investigate its dynamical properties. In particular, this scheme extends our previous elastic lattice model for simulating tightly double-folded ring polymers, where only single monomers could move~\cite{ElhamPRE2021}, by incorporating an additional tree node move akin to diffusive dynamics. We then explain how to systematically and in a fully controlled manner combine the different types of monomer and tree node moves. Finally, we introduce the systems studied: ideal non-interacting double-folded rings, isolated self-avoiding double-folded rings, and melts of double-folded rings.
In the Results Section~\ref{sec results}, we briefly examine the well-understood static ring properties of these systems, which also serve to validate that the ensembles generated by different combinations of local dynamics remain identical as long as the system conditions are unchanged.
We then focus on comparing the diffusion properties of both monomers and the center of mass across different dynamical schemes with available theoretical results. Notably, in melts, we find that the mixture of tree node and ring monomer moves leads to a non-trivial acceleration of the emergent dynamics.
The subsequent Discussion Section~\ref{sec:Discussion} explores the functional form and amplitude of the dynamics arising from the coupling in a context that deals more generally with the analysis of data from mixed dynamical schemes and the question of how to identify and distinguish non-trivial coupling effects from situations where the contributions of different local modes to the emergent dynamics are independent.
Finally, we provide a brief conclusion in Section.\ref{sec:CONCLUSION}.

%
%
%
%
\section{THEORETICAL BACKGROUND }\label{sec:theory}

\subsection{Static Properties}
The behavior of a double-folded ring polymer~\cite{Khokhlov1985Phy.L.A,Rubinstein1986PhysRevLett.,Obukhov.Rubinstein1994PRL,Halverson:2011JCP,SmerkGrosberg2015Cond.Matt,Rosa2014Phys.Rev.Lett.,MichielettoSoftMatter2016} can be analyzed at three structural levels~\cite{Rosa2019The.EPJE} (Fig.~\ref{fig:WrappedRing}). The primary structure is defined as the connectivity of the monomers along the ring. The secondary structure corresponds to the double folding. It can be represented by mapping the ring onto a graph with the connectivity of the tree. The tertiary structure is defined by the positions of the ring monomers and tree nodes within the embedding (three-dimensional) space.
A small set of exponents describes how the expectation values of typical observables scale with the ring weight $N$ (the number of monomers): 

	\begin{equation}
		\langle L (N) \rangle  \sim N^{\rho},
		\label{eqn:L}
	\end{equation}

	\begin{equation}
		\langle R^2_g (N) \rangle \sim N^{2 \nu}.
		\label{eqn:Rg}
	\end{equation}
where $L$ is the mean tree contour distance (the shortest distance on the tree) between tree nodes, $R_g$ is the radius of gyration of the ring~\cite{ElhamPRE2021}. For ideal, non-interacting systems, the exponents are $\rho=1/2$ and $\nu=1/4$~\cite{ZimmStockmayer49,DeGennes1968}. 
For interacting systems, the only known exact result is the value $\nu=1/2$ for self-avoiding trees in $d=3$~\cite{ParisiSourlasPRL1981}. 
Otherwise, Flory theory~\cite{IsaacsonLubensky1980,DaoudJoanny1981,Grosberg93Universality,Grosberg2014SoftMatter,Ralf2017SoftMatter} predicts $\rho=2/3$  for self-avoiding trees, and $\nu=1/3$ and $\rho=5/9$ for trees in the melt state.

\subsection{Dynamical Properties}\label{subsec:DynamicalProperties}
A typical analysis of polymer dynamics involves monitoring, as a function of time:
\begin{itemize}
    \item the total monomer mean-square displacement (MSD)
        \begin{equation}
            g_1(t) = \langle {|\mathbf{r}_i(t) - \mathbf{r}_i(0) |}^2 \rangle
            \label{eqn:g1}
        \end{equation}
    \item   the monomer MSD relative to the chain’s center of mass (CM)
        \begin{equation}
            g_2(t) = \langle {|\mathbf{r}_i(t) -\mathbf{r}_{CM}(t) + \mathbf{r}_{CM}(0)-\mathbf{r}_i(0) |}^2 \rangle
            \label{eqn:g2}
        \end{equation}
    \item the MSD of CM
        \begin{equation}
            g_3(t) = \langle {|\mathbf{r}_{CM}(t) - \mathbf{r}_{CM}(0) |}^2 \rangle.
            \label{eqn:g3}
        \end{equation}
\end{itemize}
On time scales where the rings have moved beyond their own size, their CM exhibit diffusive motion: $g_1(t) = g_3(t) \sim t$, and $g_2(t) = 2 \langle R^2_g\rangle$.
On shorter time scales, the internal dynamics can display a range of intermediate scaling regimes, whose understanding is important to comprehend the diffusion constant in the asymptotic regime. A summary of the scaling exponents for $g_1(t)$, $g_3(t)$, and the diffusion constant $D$ in the known limiting cases are discussed below and given in Table~\ref{tab:exponents}.

\subsubsection{Slithering dynamics of tightly wrapped rings around fixed trees}\label{subsubsec:Slitheringdynamics}

For local mass transport via reptons along a fixed tree~\cite{DeGennes1971,DoiEdwards,RubinsteinRepton1987,MarkoSiggia1995,MilnerPhysRevLett.2016}, the monomer motion  {\it in the embedding space} is given by:

\begin{equation}
g_1(t)\sim \langle \delta s^2(t)\rangle^\nu \ .
 \end{equation}
where $\langle \delta s^2(t) \rangle$ denotes the  one-dimensional monomer MSD {\it along the tree}. Several regimes must then be distinguished. First, 
for the shortest time scales, $\langle \delta s^2(t) \rangle \sim t$, such that:
\begin{equation}
g_1(t) \sim t^{\nu} \ .
 \end{equation}
At longer time scales, the ring connectivity gives rise to one-dimensional Rouse dynamics typical of a linear polymer, $\langle \delta s^2(t) \rangle \sim t^{1/2}$, such that:
\begin{equation}
g_1(t)  \sim  t^{\frac{\nu}{2}} \ .
\end{equation}
Beyond the Rouse time, $\tau_R \sim N^2$ where $g_1(\tau_R) \sim N^\nu \sim \sqrt{\langle R_g^2\rangle}$,
 the dynamics crosses over to the second reptation regime where all monomers diffuse coherently around the tree. Again $\langle \delta s^2(t) \rangle \sim t$, so that: 

 \begin{equation}
g_1(t)\sim t^{\nu},
 \end{equation}
 but, compared to the shortest time scales, with a $N$ times smaller diffusion constant of the one-dimensional CM motion.
Finally, as the ring diffuses along a fixed, closed path, the monomer displacements reach a plateau at $2\langle R_g^2\rangle$.
$g_3(t) \ll \langle R_g^2\rangle$, because fluctuations in the ring CM are only due to negligible density fluctuations along a ring, which is stretched over a path whose length is of the order of the ring length. 

\subsubsection{Rings in an array of fixed obstacles}\label{subsubsec:RIAO}

The combination of repton moves with the local creation or retraction of side branches gives rise to the characteristic motion of rings in an array of fixed obstacles~\cite{Rubinstein1986PhysRevLett.,SmerkGrosberg2015Cond.Matt} suppressing the sideways motion of tree nodes. 
Specifically,  $g_1(t)\sim t^{\frac{2\nu}{2+\rho}}$ and $g_3(t)\sim t^{\frac{2\nu+1}{2+\rho}}$ in the intermediate regime~\cite{SmerkGrosberg2015Cond.Matt}. Moreover, the CM diffusion coefficient is found to scale as $D\sim N^{-2-\rho+2\nu}$~\cite{SmerkGrosberg2015Cond.Matt}.

\begin{table}
\caption{
Summary of the scaling exponents for both static and dynamic properties of ideal rings, isolated self-avoiding (S.A.) rings, and self-avoiding rings in the melt state. Specifically, the table reports the dynamic exponents for $g_1(t)$, $g_3(t)$, and the diffusion constant $D$, as well as the static exponents $\nu$ and $\rho$. }
\label{tab:exponents} 
\setlength{\tabcolsep}{3.5pt}
\renewcommand{\arraystretch}{1.4}
  \begin{tabular}{ l lc c  c c cl l }
    \multirow{1}{*}{} & &$\nu$ & $\rho$ & $g_1$ & $g_3$ & $D$ \\
        \hline
        & Ideal rings  & $1/4$ &$1/2$\\
        \hline
        & Rouse Dynamics for the quenched tree &  &  &$1/3$ & $1$ & $-1$ \\
  	& Ring in an array of obstacles &  & &$1/5$ & $3/5$ & $-2$\\
        & Ring slithering around the tree &  &  &$ 1/4,1/8,1/4$ & $1/4$&  $0$ \\
	 \hline
	 & S.A. rings  &$1/2$ & $2/3$\\
        \hline
        & Rouse Dynamics for the quenched tree&  & &$ 1/2$ & $1$& $-1$ \\
        & Ring in an array of obstacles&  & &$3/8$ & $3/4$& $-5/3$ \\
  	& Ring slithering around the tree& &  & $1/2,1/4,1/2$ & $1/2$ &$0$ \\
	 \hline
        & Melt rings  &$1/3$ & $5/9$\\
        \hline
        & Rouse Dynamics for the quenched tree&&  & $ 2/5$ & $1$& $-1$ \\
        & Ring in an array of obstacles&  & &$6/23$ & $15/23$& $-17/9$ \\
  	&  Ring slithering around the tree &&  & $1/3,1/6,1/3$ & $1/3$ &$0$ \\
    &  Fractal Loopy Globule (FLG) &  &  &$2/7$ & $5/7$ &$-5/3$ \\
    \end{tabular}
\end{table}

\subsubsection{Brownian dynamics of randomly quenched trees}\label{subsubsec:Qtree} 

The emerging dynamics of quenched trees, where only tree nodes can diffuse via Brownian motion, are expected to be described by a suitably generalized Rouse model~\cite{DeGennesBook} predicting free diffusion of the tree CM at all time scales, $g_3(t)\sim t/N$, $g_1(t)\sim t^{\frac{2\nu}{2\nu+1}}$ in the intermediate regime and $g_1(t)\sim t$ on the shortest time scales. Moreover, the CM diffusion coefficient is expected to scale as  $D\sim N^{-1}$.


\subsubsection{Rings in an array of transient obstacles}\label{subsubsec:FLG} 

The dynamics in the melt state is subject to local topological constraints due to the mutual impenetrability of double-folded ring sections. However, to the extent that the rings move, these constraints are not fixed, but transient. 
The \textit{fractal loopy globule} (FLG) model incorporates the effect of moving obstacles on the dynamics of the target ring~\cite{Rubinstein:Macromolecules2016}. According to this model, topological constraints in the melt state gradually dilute over time. This occurs because, as time progresses, larger loops relax, eventually no longer acting as obstacles. This process is analogous to the tube dilation phenomenon observed in polydisperse linear melts~\cite{McLeish:2002AdvancesinPhysics,Gold:2019Phys.Rev.Lett.,Colmenero:2019PhysRevLett.}. The FLG model predicts the scaling of the mean square displacement as $g_1(t)\sim t^{\frac{2\nu}{2+\nu}}$, and the diffusion coefficient scaling as $D\sim N^{-2+\nu}$.

%
%
%
%
\section{MODEL AND METHOD}\label{sec:Model and Method}

\subsection{Discrete Double-Folded Ring Model}\label{subsec:ring}

We consider discrete rings made of $N_m$ monomers embedded in a face-centered cubic (FCC) lattice, with the mesh size (i.e., the space unit) set to $1$. Following previous works~\cite{Obukhov.Rubinstein1994PRL,ElhamPRE2021} and the original elastic lattice polymer models~\cite{EvansEdwards1981-Part1,Barkema.Book1999,Barkema2003The.JofCh.Phy,Kolb2009Macromolecules}, we include the possibility that two successive monomers along the ring may occupy the same lattice site. In this case, the bond length between these monomers is $0$ (referred to as stored length), compared to $1$ for extended bonds (Fig.~\ref{fig:WrappedRing}a). The double-folded nature of the model is then imposed by pairing each extended bond of the double-folded polymer with an opposing extended bond, as shown in Fig.~\ref{fig:WrappedRing}a.

\subsection{Handling the Underlying Lattice Tree}\label{subsec:tree}

A randomly double-folded ring polymer is characterized by an underlying randomly branched tree structure~\cite{Khokhlov1985Phy.L.A,Rubinstein1986PhysRevLett.,Obukhov.Rubinstein1994PRL,Halverson:2011JCP,SmerkGrosberg2015Cond.Matt,Rosa2014Phys.Rev.Lett.,MichielettoSoftMatter2016}. This tree is an undirected graph, whose nodes (vertices of the graph) are located on the occupied lattice sites and edges correspond to double-folded ring bonds. By definition, the tree is acyclic, and any pair of tree nodes is connected through a unique shortest path on the tree.  
Fig.~\ref{fig:WrappedRing} provides a visual $2D$ representation of a double-folded ring together with its underlying tree structure.

In our previous work~\cite{ElhamPRE2021}, we simulated double-folded ring polymers and characterized the corresponding tree conformations only during post-processing. This is unproblematic in systems with hard excluded volume interactions, where a ring configuration corresponds to a unique tree structure, because all spatially co-localized monomers belong to the same tree node (Fig.~\ref{fig:WrappedRing}). However, in ideal double-folded rings, multiple tree nodes may occupy the same lattice site, introducing ambiguity in defining the underlying tree structure~\cite{ElhamPRE2021} and the statistical weight of ring conformations~\cite{Pieter:PhysRevE.2024}

To solve this problem, we extend the description of the double-folded ring by considering the degrees of freedom associated with the underlying tree. To this end, we define a ring conformation ${(\cal C)}$ by its set of ($N_m$) monomer positions in the embedding space, $\{\vec{r}_1,\vec{r}_2, ..., \vec{r}_{N_{m}}\}$, together with its set of ($N_n$) tree node positions, $\{\vec{R}_1,\vec{R}_2, ..., \vec{R}_{N_n}\}$, the ring and tree connectivity graphs, and information about which monomers belong to which tree node. For ideal systems we impose the constraint that nearest-neighbor tree nodes must not overlap, i.e.~$|\vec{R_I}-\vec{R_{J}}|=1$, which is automatically fulfilled in systems with hard excluded volume interactions. 
As a consequence, there are small differences in the local structure of the present ideal rings and those of \cite{ElhamPRE2021}. A summary of the data structure is presented in Table~\ref{table:notation}.

\begin{table}[b!]
\fontsize{8}{10}\selectfont
\renewcommand{\arraystretch}{1.8}
\setlength{\tabcolsep}{1pt}
\centering
  \begin{tabular}{|c | c| c| } 
 \hline
  & Ring Monomer $i$ & Tree Node $I$  \\ [0.5ex] 
 \hline\hline
Position & $\overrightarrow{r_i}$ & $\overrightarrow{R_I}$  \\ 

 Functionality                & $2$                                         & $f_I$  \\ 
 
Connectivity                 & $\Gamma(i)= \{i-1,i+1\}$                    & $ \Gamma(I)= \{J,K,...\}$  \\
			                & $|\Gamma(i)|= 2$                    & $ |\Gamma(I)|= f_I$  \\

Bond Constraints            &     $|\vec{r_i}-\vec{r_{i+1}}|\in\{0,1\}$     &  $|\vec{R_I}-\vec{R_{J}}|=1$  \ $\forall J \in  \Gamma(I)$  \\[1ex]

Ring $\leftrightarrow$ Tree & $i \rightarrow I$                                        & $I \rightarrow \{i,j,k,...\}$   \\

 \hline
  \end{tabular}
 \caption{ Data structure defining the conformation ${(\cal C)}$ of a double folded ring polymer and the associated tightly wrapped tree. 
 }
 \label{table:notation}
\end{table}

In this context, the functionality $f$ of a tree node (the degree of a vertex) is defined by the number of tree edges coming out from the node. 
Specifically, a node with $f=1$ corresponds to a leaf or branch tip, $f=2$ to a linear section of the double-folded ring, and $f > 2$ to a branching point. 
For simplicity, and without loss of generality~\cite{Rosa2014Phys.Rev.Lett.}, we limit the functionality to $f_{\text{max}}=3$ in this work (For a more detailed analysis of this choice, please see Sec.~II in the supplementary material.).

\subsection{Dynamic Monte Carlo Scheme for Mixed Dynamics}
\label{sec:kMC}

To investigate the dynamical properties of double-folded rings, we performed dynamic MC simulations. To this end, we used two types of monomer moves and one new type of tree node move (see first subsection below), together capturing the three possible types of local dynamics in double-folded polymers: reptons, creation/retraction of side branches, and Brownian motion of tree nodes. In addition, when considering volume exclusion (see systems studied in Section~\ref{sec:systems}), we employed a Metropolis algorithm, setting a very high interaction energy (i.e., greater than $30$ units of thermal energy) between nodes on the same lattice site, with no interaction otherwise.

As explained in the second subsection below, we then combined these three types of trial moves to generate mixed dynamics in a controlled manner, ensuring that the time unit reflects the fraction of time each type of move is selected.

\subsubsection{Trial moves}
\label{sec:trial}

Trial moves consist of individual monomers or tree nodes attempting to hop to adjacent lattice sites, with three distinct types:
\begin{description}
\item[Repton (monomer) move] This occurs when a monomer hops longitudinally along the ring, crossing an extended bond to a neighboring tree node, thereby transporting a unit of stored length along the tree without changing its structure or the system's energy. An example is shown in Fig.~\ref{fig:MC_moves}a. A detailed analysis of the repton distribution is provided in the Supplementary Material, Sec.~III.

\item[Hairpin (monomer) move] This occurs when a monomer flanked by two zero-length bonds hops in a transverse direction, extending or creating a side branch. During the move, both zero-length bonds unfold, generating two new paired, oppositely oriented extended bonds. Consequently, the moved monomer is reassigned to a new node of functionality $f=1$. This new node connects to the node to which the monomer was originally associated, increasing the latter's functionality by one, as well as the total number of tree nodes, $N_n$. Consequently, $N_n$ varies during a simulation involving hairpin moves. The corresponding inverse move removes an extended bond pair, leading to the shortening or complete removal (annihilation) of a side branch. An example is shown in Fig.~\ref{fig:MC_moves}b. 

\item[Tree node move] This occurs when a tree node moves with all its associated monomers without changing the length of any of the involved tree edges and ring bonds. These moves thus preserve the tree structure and the association between the tree and the ring. Examples are shown in Fig.~\ref{fig:MC_moves}c-d.

\end{description}

\begin{figure}[H]
    \centering
    \includegraphics[width=\columnwidth]{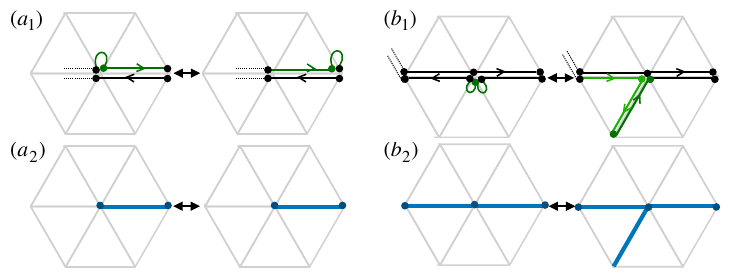}%
    \qquad
    \includegraphics[width=\columnwidth]{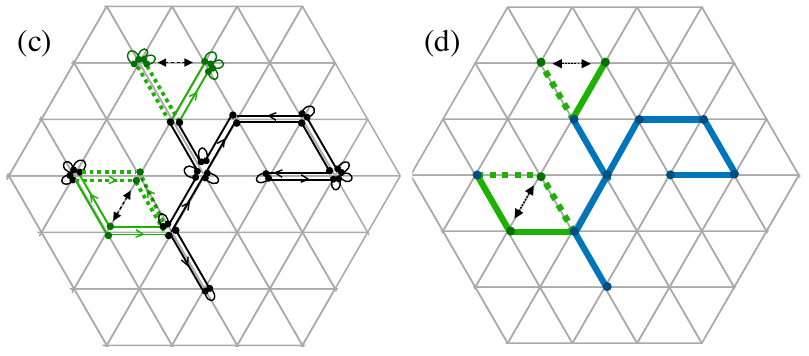}%
    \caption{(a)-(b):A monomer move involves hopping a randomly selected monomer into the nearest neighbor lattice site. (a) The Repton move transports stored length along the ring and locally redistributes it without altering the tree structure. (b) The Hairpin move is responsible for the creation or annihilation of side branches. The top row shows the double-folded ring, and the bottom row shows the underlying tree structure. (c)-(d) Two examples of tree node trial moves, where a randomly selected tree node attempts to move to one of the twelve nearest neighbors of the face-centered cubic (FCC) lattice used in our simulations, subject to a double-folding preserving constraint. For clarity, a two-dimensional schematic is shown here, while the actual simulations are performed in three dimensions.}%
    \label{fig:MC_moves}
\end{figure}

\subsection{Mixed Dynamics and the Unit of Time}
\label{sec:mixed_method}


In dynamic MC simulations, the (effective) time unit $\tau$ is equivalent to one MC sweep comprising a number of $M_s$ MC trials or steps chosen such that each component of the system is selected on average once. For instance, if the dynamics consists solely of repton or hairpin moves, a sweep comprises $M_s=N_m$ trials, associating a duration of $\tau_m = \tau/N_m$ with an individual MC step. If the dynamics consists solely of tree node moves, a sweep instead consists of $M_s=N_n$ trials and the time associated with a step is $\tau_n = \tau/N_n$. In particular, $\tau_n/\tau_m = N_m/N_n > 1$ because a node move displaces several associated monomers.

A natural way to define a mixed dynamics is to randomly alternate monomer and tree node {\em sweeps} with probabilities $p_m \in [0,1]$ and $(1-p_m)$ respectively. 
For $p_m=1$ this essentially generates the rings-in-an-array-of-obstacles dynamics from Ref.~\cite{ElhamPRE2021}, while $p_m=0$ simulates the Brownian dynamics of quenched trees. Intermediate values mix the two types of motion in arbitrary proportions. 

In the above scheme a total of $t$ sweeps corresponds to $p_m N_m t$ monomer moves and $(1-p_m) N_n t$ node moves. In practice, we have used the following scheme to alternate between monomer and node {\em steps} in such a way as to preserve the above proportions: 
\begin{enumerate}
    \item Compute the number $M_s$ of MC steps for the next sweep as:
    \begin{equation}
        M_s = p_m N_m + (1 - p_m) N_n
        \label{eqn:MCsweep}
    \end{equation}
    Note that since $N_n$ can vary, $M_s$ can also vary, but that irrespective of the value of $N_n$ moving all tree nodes always corresponds to moving all monomers.
  	 \item Perform \(M_s\) trials. For each trial, randomly select a move type: a monomer move with probability \(p_m N_m / M_s\), or a tree node move with probability \((1 - p_m) N_n / M_s\).  
	Since the move type is drawn independently for each trial, this procedure amounts to sampling with replacement. The number of monomer and node moves fluctuates from sweep to sweep, but their statistical proportions are satisfied on average.
	\item Increment time by \(1\).

\end{enumerate}
By construction, $p_m N_m \tau_m + (1 - p_m) N_n \tau_n = \tau$ and a monomer is moved on average once per sweep. 

In addition, we have introduced acceptance probabilities $0 \le q_{rep},q_{hp} \le 1$ for the two monomer moves.
For $q_{rep}=0$ and $q_{hp}=1$, all Repton moves are suppressed, resulting in a dynamics that involve only the Hairpin moves. In the opposite case of $q_{rep}=1$ and $q_{hp}=0$, the only accepted monomer moves are Repton moves generating mass transport along the tree.
While equal values  $0 \le q_{rep} = q_{hp} < 1$  generate a slowed down version of the original dynamics, other choices allow to arbitrarily manipulate the relative weight of Repton and Hairpin moves.

\subsection{Systems Studied and Equilibrium Preparation}
\label{sec:systems}

We performed simulations for double-folded rings composed of $N_{m} =64, 125, 216,343, 512, 1000$ monomers on a three-dimensional FCC lattice with periodic boundary conditions for three types of systems: ideal rings, isolated self-avoiding rings, and rings in the melt state. 
For the ideal rings, there were no restrictions on the number of tree nodes per lattice site (i.e., no excluded volume interaction). 
For the self-avoiding rings and the rings in the melt state, we included excluded volume interactions (explained in Section~\ref{sec:kMC}). Self-avoiding rings were simulated in a sufficiently large box, while a high lattice occupation number, $\phi \simeq 0.93$, was used for the melt state. A summary of the simulation parameters and equilibrium values for the studied systems is provided in Table~I in Supplementary Material.

All reported results were generated by first fully equilibrating the system according to the protocol of~\cite{ElhamPRE2021}, using monomer moves for annealed double-folding rings (see Supplementary Material, Fig.~S1 in Section~I). Then, at the initial time of all simulation runs, an ``equilibrated'' state was randomly chosen. 

%
%

\section{RESULTS}\label{sec results}
To set the stage, we begin with a brief overview of the systems' static properties in Section~\ref{subsec:Statistics}, which also serves to validate our simulation methods. In Sections~\ref{sec:Dynamic results}~and~\ref{sec:scaling}, we analyze the emergent dynamics of monomers and CM for different combinations of trial moves.

\subsection{Static Properties }\label{subsec:Statistics}
 \begin{figure}[h]
\centering
\includegraphics[width=\columnwidth]{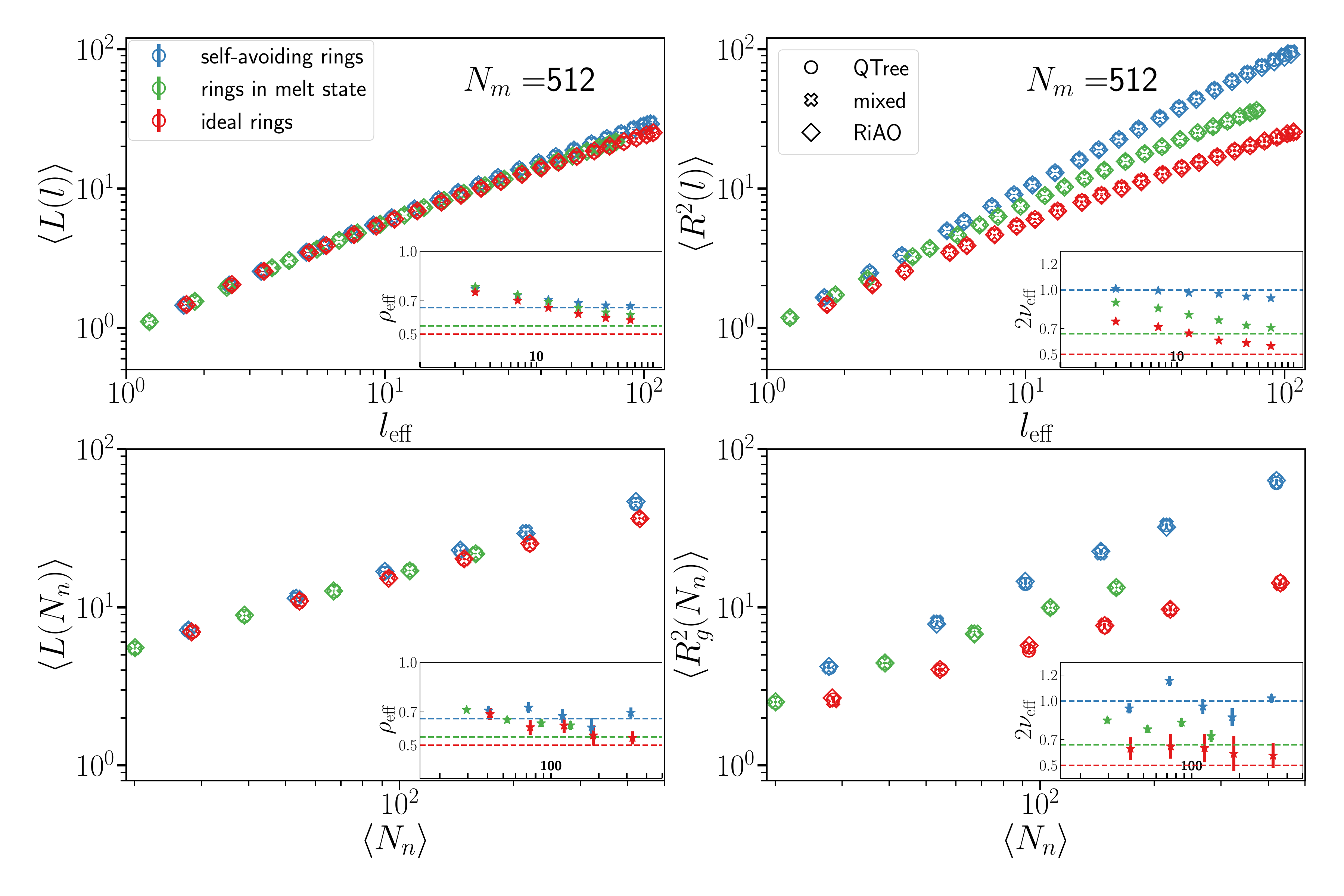}
\caption{ Conformational statistics for isolated ideal rings, isolated self-avoiding rings, and rings in the melt state when subjected to  Quenched Tree dynamics (QTree), ring in an array of obstacle-like (RiAO) dynamics, and the combination of the two (mixed, with $p_m=0.5$). Insets show the local slopes of the data; horizontal dashed lines correspond to the expectation scaling exponents for each system (summarized in Table~\ref{tab:exponents}). 
The conformational statistics should be independent of the dynamics if the system condition remains the same and the results in the figure strongly confirm this assertion. Error bars are the same size or smaller than the symbols. }\label{fig:stat_Q_A}
\end{figure}

In Figure~\ref{fig:stat_Q_A} we present a number of results characterizing the secondary structure and the tertiary structure of our double folded rings along the lines of Ref.~\cite{Rosa2019The.EPJE,ElhamPRE2021}. Different colors distinguish results for ideal, self-avoiding and melt systems, while the symbols indicate the type of dynamics we used to acquire the data. 

Specifically we have calculated for all pairs $(i,j)$ of ring monomers their distance, $L(i, j)$, on the tree and their square distance, $R^2(i,j)$, in space. 
Averages as a function of the ring contour distance, $l = |i-j|\langle b \rangle$ are plotted in the top row of Fig.~\ref{fig:stat_Q_A}. The average bond length $\langle b \rangle$ is smaller than one, since the monomer bonds can be $0$ or $1$ due to the presence of reptons in our model.
Note that the data shown in the top row of Figure~\ref{fig:stat_Q_A} is plotted as a function of the effective ring contour distance, $l_\text{eff} = l(1- {|i-j|}/N_m)$, as a means of reducing finite ring size effects~\cite{Rosa2019The.EPJE}.
As a complement, we averaged $L(i, j)$ and $R^2(i,j)$ over all monomer pairs of our rings to calculate the average tree contour distance $\langle L\rangle$ and the mean-square gyration radius $\langle R_g^2\rangle$. Results as a function of the average number of tree nodes, $\langle N_n\rangle$, are plotted in the panels in the bottom row of Fig.~\ref{fig:stat_Q_A}. A summary of these values for the studied systems is provided in Table~I in the Supplementary Material. Additionally, we calculated the effective local exponents, $\rho_{eff}$ and $\nu_{eff}$, from the slopes of the data points shown in the four panels of Fig.~\ref{fig:stat_Q_A}. The results are presented in the insets of the corresponding panels.

In absolute terms, the differences between the average tree contour distances,  $\langle L(l_\text{eff}) \rangle$ and $\langle L(N_n) \rangle$, for the three ring types are small. Nevertheless our data are compatible with the expected crossover from local linear behavior with $\langle L(l) \rangle = l$ to $\langle L(l) \rangle \sim l^\rho$ with $\rho=1/2$ for ideal trees, $\rho=2/3$ for self-avoiding trees, and $\rho=5/9$ for trees in the melt state.

In contrast, there are marked differences between the mean square spatial distance between ring monomers, $\langle R^2(l_\text{eff})\rangle$ and $\langle R^2(N_n)\rangle$, for the different ring types. 
For ideal trees, this quantity is given by~\cite{Rosa2016JofPh.A} $\langle R^2(l)\rangle = l_K \langle L(l) \rangle$, where the Kuhn length $l_K$ for the present model is simply given by the lattice constant. The fact that the independently calculated data sets for $\langle L(l_\text{eff}) \rangle$ and $\langle R^2(l_\text{eff})\rangle$ look indistinguishable presents a simple and independent check of the path analysis on the tree.
As expected, self-avoiding rings are more strongly swollen relative to ideal rings than rings in a melt. 
Again, our data are compatible with the expected crossover from local linear behavior with $\langle R^2(l) \rangle = l$ to $\langle R^2(l) \rangle \sim l^{2\nu}$ with $\nu=1/4$ for ideal trees, $\nu=1/2$ for self-avoiding trees, and $\nu=1/3$ for trees in the melt state.
Note, however, the deviation of the effective exponents on small scales from the expected asymptotic values, which suggests the existence of corresponding deviations for effective dynamic exponents.

Importantly, for the present purpose, there are no noticeable differences between the data acquired from different dynamical schemes. Figure~\ref{fig:stat_Q_A} thus validates our simulation methods and confirms that starting from an ensemble of equilibrated rings all ergodic dynamical schemes generate the same conformational statistics independently of whether the tree connectivity is randomly quenched or annealed. 

%
%
\subsection{Overview of the Emergent Dynamics}\label{sec:Dynamic results}

\begin{figure}[H]
\includegraphics[width=\textwidth]{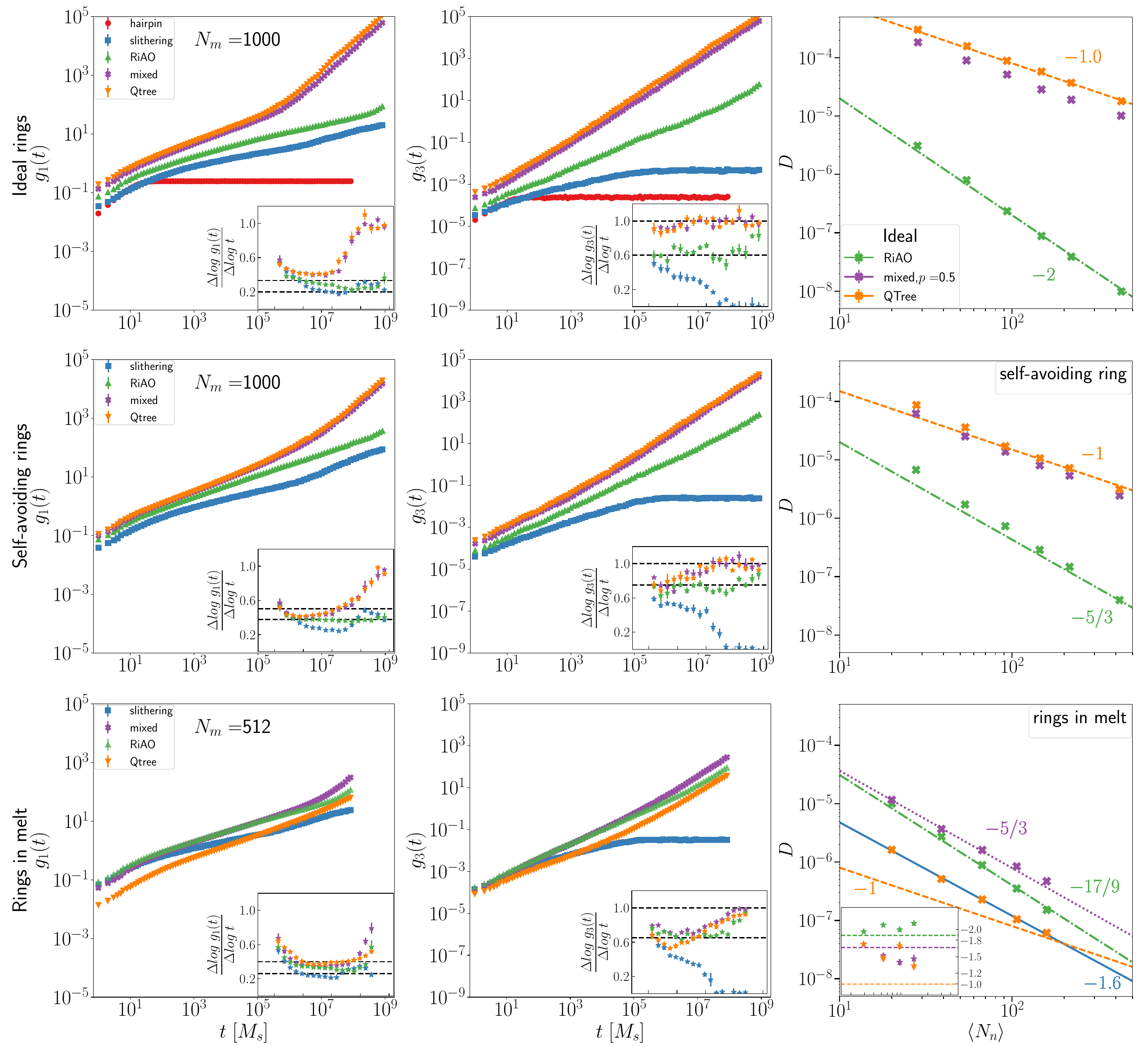}
\caption{Overview of emergent dynamics for different ring polymer systems: ideal (top row), isolated self-avoiding (middle row), and double-folded ring polymers in melts (bottom row). Results are shown for: monomer MSD, $g_1(t)$ (left column), CM MSD , $g_3(t)$, (middle column) and the size dependence of the asymptotic CM diffusion constants, $D(N_n)$, (right column). Dynamics considered include: hairpin move, slithering dynamics, ring-in-an-array-of-obstacles (RiAO)-like dynamics, quenched tree dynamics (QTree), and a mix of RiAO and QTree dynamics with $p_m=0.5$. Insets show the local slopes of the data, with dashed lines indicating the expected exponents for QTree and RiAO dynamics (summarized in Table~\ref{tab:exponents}). Error bars are smaller than or comparable to the symbol sizes.}\label{fig:OverviewDynamics}
\end{figure}

Figure~\ref{fig:OverviewDynamics} presents an overview of the emergent dynamics for three classes of double-folded rings -- ideal rings (top row), isolated self-avoiding rings (middle row), and ring polymers in melts (bottom row) -- for which we have investigated the impact of the following local dynamics:
\begin{enumerate}
    \item the exclusive application of hairpin moves,
    \item a slithering dynamics around fixed trees as a result of the exclusive application of repton moves,
    \item a RiAO dynamics as a result of the combined application of the two monomer moves,
    \item a Brownian dynamics of trees with quenched connectivity (QTree) as a result of the exclusive application of tree node moves,
    \item the combination of ring monomer and tree node moves, using $p_m = 0.5$.
\end{enumerate}
Results for the monomer MSD, $g_1(t)$, the CM MSD, $g_3(t)$, and the size dependence of the asymptotic CM diffusion constants $D = \lim_{t\to\infty} \frac{g_3(t)}{6t}$, are presented in separate panels in the three columns of the figure. In addition, we report results obtained for the combination of repton and tree node moves in Fig.~S6 in sec.~V in the supplementary material. 


While hairpin moves can locally redistribute accumulations of stored length, the overall dynamics emerging from their exclusive application is close to negligible: the final plateau is reached after approximately $\sim30 M_s$, with monomer and CM MSDs remaining strictly proportional to each other, i.e., $g_1(t)=N_m g_3(t)$. In contrast, the slithering dynamics emerging from the exclusive application of repton moves allows ring monomers to explore, in the long run, all sections of a wrapped, fixed tree. Notably, the confinement of the motion becomes visible earlier for the ring's CM than for individual monomers. Importantly, it is only the combination of the two monomer moves that makes the RiAO dynamics ergodic. Interestingly, the absolute magnitude of the latter varies only weakly between the three considered systems.

For ideal and isolated self-avoiding rings, the quenched tree (QTree) dynamics is significantly faster than the RiAO dynamics, because in the intermediate scaling regime the exponents describing the (in general anomalous) monomer and CM diffusion are effectively larger. As a consequence, the rings equilibrate more quickly as it takes them less time to
displace over distances comparable to their own size. 

The situation is quite different in the melt case, where the hard-excluded volume interactions and the high occupation number ($\phi \sim 0.93$) of the lattice cells induce a strong reduction in the acceptance probability of hairpin and tree node moves, 
thus affecting both RiAO and QTree dynamics.
As the slithering component remains active, it dominates the early-time monomer motion in RiAO dynamics. Consequently, at short times RiAO dynamics exhibits faster monomer motion than QTree dynamics.
However, in the intermediate-time regime QTree dynamics shows slightly larger effective exponents than RiAO  (inset in bottom left panel in Fig.~\ref{fig:OverviewDynamics}). Overall, for the present ring sizes RiAO  is faster than QTree dynamics. However, our data for the size dependence of the diffusion constant, $D(N_n)$, (bottom right panel in Fig.~\ref{fig:OverviewDynamics}) suggest that QTree dynamics is asymptotically faster than RiAO dynamics.


For ideal and isolated self-avoiding rings, the MSD resulting from the combination of ring monomer and tree node moves is unremarkable, displaying an intermediate behavior between the MSD emerging from the exclusive application of either monomer moves or node moves. Namely, with our definition of a MC sweep as the unit of time (see Section~\ref{sec:mixed_method}
), the mixing of 50\% tree node moves and 50\% of the much less effective monomer moves essentially slows down the motion by a factor of two in time. 
By contrast, and notably, something interesting happens in the melt case: \textit {the mixed dynamics is faster than either of the pure components}, what we investigate in detail below.

\subsection{Scaling analysis of the Emergent Dynamics}\label{sec:scaling}

Before analyzing in detail the speed-up associated with the combined dynamics, we first present a detailed scaling analysis of the dynamics reported in Fig.~\ref{fig:OverviewDynamics}. This analysis begins with the three limiting cases discussed in the Theoretical Background (Section~\ref{sec:theory}) and concludes with the mixed dynamics.

Specifically, in Fig.~\ref{fig:Slithering_scaling} we report scaling results for $g_1$ and $g_3$ in the context of the exclusive application of repton moves (slithering dynamics), for ideal trees. In Fig.~\ref{fig:RAO_BQT_Mix_scaling}, we report corresponding scaling results in the context of three other dynamics: the RiAO dyanmics (left column), the Brownian motion of trees (central column) and the combined dynamics (right column), for ideal rings (top row), self-avoiding rings (middle row), and rings in melts (bottom row). Additional details and figures are provided in Figs. S9–S15 of Sections~VIII–IX in the Supplementary Material.

\subsubsection{Slithering dynamics around fixed ideal trees}\label{subsec:slitheringideal}
Let us first consider the slithering dynamics around a fixed (quenched) tree emerging from the exclusive application of repton moves. 
In Fig.~\ref{fig:Slithering_scaling} (see also Fig.~S15 in the Supplementary Material for more detailed results), we show the corresponding monomer and CM MSD for ideal randomly branching double-folded rings composed of $N_m=64,216,512,1000$ monomers. An obvious feature is the nearly absent CM motion of the rings, $g_3(t) < \langle R_g^2(N_m) \rangle / N_m$, such that $g_1(t) \approx g_2(t)$. We then use various rescaling of the same data in the different panels of the figure to reveal different time and length scales relevant to the monomer and the CM motion. The results are shown in Fig.~\ref{fig:Slithering_scaling}. Namely, the ring monomers are fully mobile while remaining bound to the quenched tree conformations that the rings are wrapping. As a consequence, $\lim_{t\rightarrow\infty} g_2(t) = 2 \langle R_g^2(N_m) \rangle$. 
The asymptotic plateau is reached on the reptation time scale, $\tau_{rep} \sim N_m^3$, of the {\em rings}, while the crossover from the $t^{\nu/2}$ to the $t^\nu$ regime in monomer diffusion occurs at their Rouse time, $\tau_R \sim N_m^2$.

The CM of the rings essentially coincide with the CM of their wrapped tree. The observable small deviations are due to monomer density fluctuations along the ring contour and equilibrate over the Rouse time, $\tau_R \sim N_m^2$, of the {\em rings}.
The asymptotic magnitude of the fluctuations are expected to be $g_3(t\gg \tau_R)=2\frac{N_{rep}}{N^2_m}\langle R_g^2 \rangle$.

\begin{figure}[h!]
\centering
\includegraphics[width=\columnwidth]{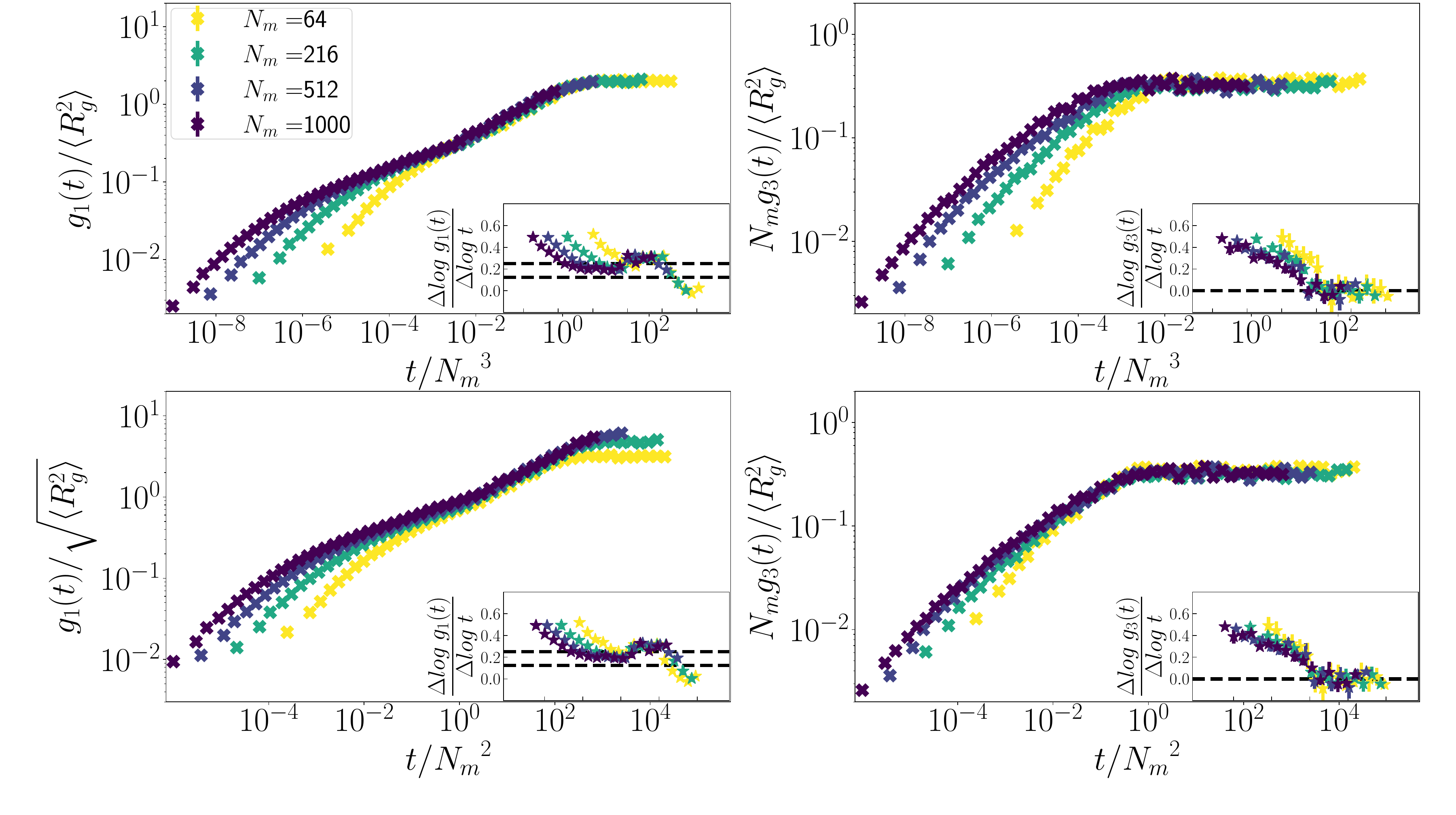}
\caption{Scaling analysis of the slithering dynamics of ideal, randomly branching, double-folded rings emerging from the exclusive application of repton moves in our model. The inset graphs in each panel share the same x-tick positions as the main panel they accompany. }
\label{fig:Slithering_scaling}
\end{figure}

\begin{figure}[h]
\centering
\includegraphics[width=\textwidth]{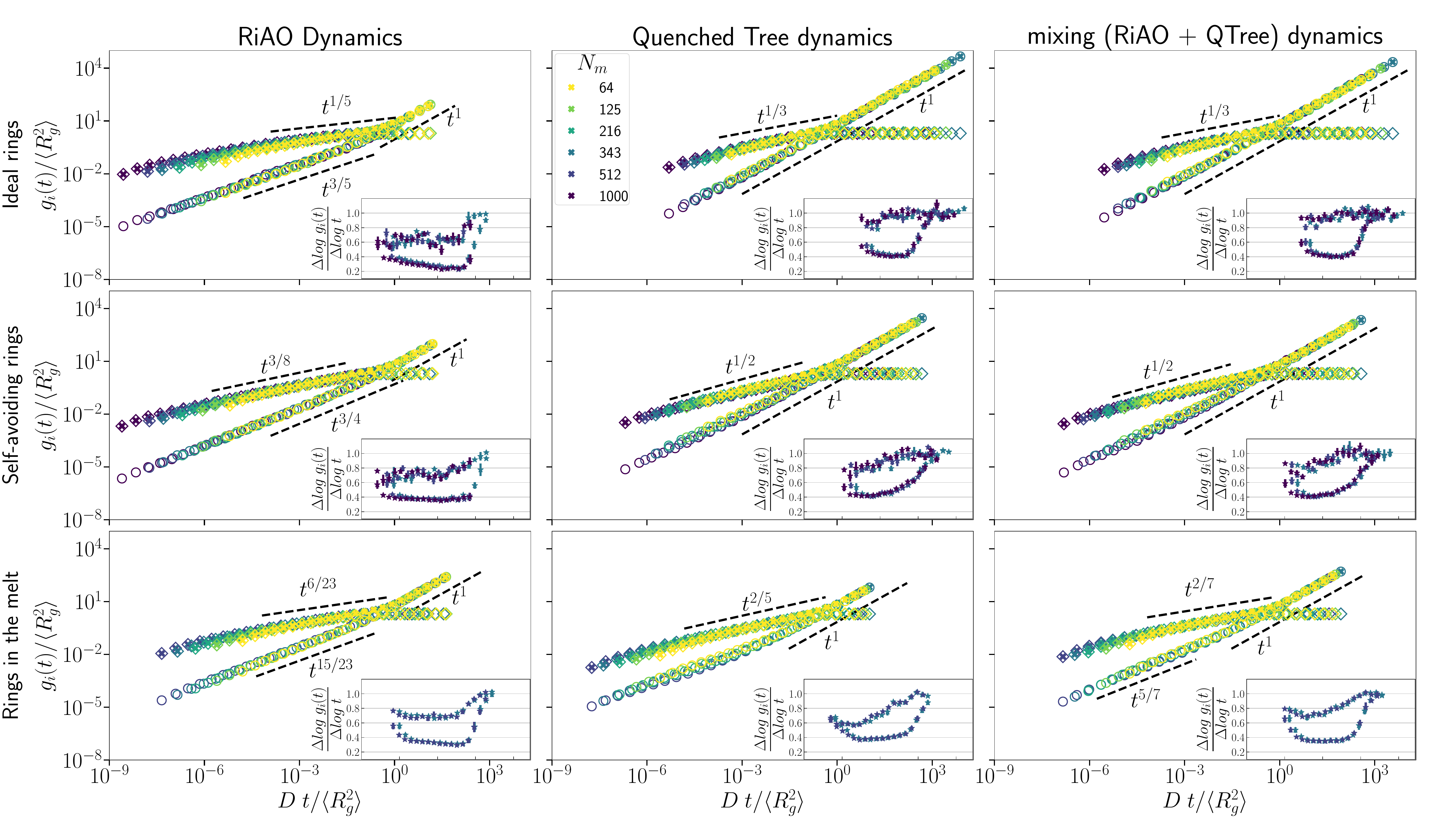}
\caption{Scaling behavior for $g_1(t)$ (crosses) $g_2(t)$ (diamonds) and $g_3(t)$ (circles) for three systems: ideal (top row), isolated self-avoiding (middle row), and double-folded ring polymers in melts (bottom row). Time is rescaled by the relaxation time, defined as $\tau_{\text{max}}=\frac{\langle R_g^2 \rangle}{D}$. Left column: Results for RiAO-like dynamics. Middle column: Results for Brownian dynamics of randomly quenched trees. Right column: Results for a mixed of the two dynamics, with $p_m=0.5$ in Eq.~\ref{eqn:MCsweep}.}\label{fig:RAO_BQT_Mix_scaling}
\end{figure}
\subsubsection{RiAO-like dynamics}\label{subsubresult:RiAO}

As already discussed in detail in Ref.~\cite{ElhamPRE2021}, the emerging dynamics resulting from a combination of repton and hairpin moves agrees with the theoretical predictions for the RiAO model (Refs.~\cite{Rubinstein1986PhysRevLett.,SmerkGrosberg2015Cond.Matt}  and Sec.~\ref{sec:theory}). As shown in the left column of Fig.~\ref{fig:RAO_BQT_Mix_scaling}, an appropriate rescaling of our present data leads again to  good agreement with the predicted $g_1(t)\sim t^{\frac{2\nu}{2+\rho}}$ and $g_3(t)\sim t^{\frac{2\nu+1}{2+\rho}}$ for the monomer and tree CM dynamics (Detailed analysis is provided in Figs.~S12--S14 in Sec.~IX of the Supplementary Material).
The corresponding effective dynamic exponents are shown in the insets of the panels. 
At least qualitatively, their slow decay in the tree scaling regime agrees with what one would expect from estimating the dynamic exponents from the effective static exponents, $\rho_{eff}$ and $\nu_{eff}$, shown in Fig.~\ref{fig:stat_Q_A}. As shown in the right column of Fig.~\ref{fig:OverviewDynamics} (green crosses), there is also good agreement between the predicted and the observed ring weight dependence of the asymptotic CM diffusion constants.

\subsubsection{Brownian dynamics of randomly quenched trees}\label{subsubresult:tree}

The emerging dynamics resulting from the exclusive application of tree node moves to our double-folded rings is presented in the central column of Fig.~\ref{fig:RAO_BQT_Mix_scaling}.  (Detailed analysis is provided in Figs.~S9--S11 in Sec.~VIII of the Supplementary Material).

For ideal and self-avoiding trees, our data are in good agreement with the predictions of the Rouse model (Tabel~\ref{tab:exponents}). The ring CM essentially exhibits simple diffusion, $g_3(t) \sim t/N_m$, with a diffusion constant inversely proportional to the number of ring monomers. A slowdown occurs independently of the ring size only over the first $50$ MC sweeps in the ideal trees, while in the self-avoiding case, the slowdown persists for a longer duration, up to $1000$ sweeps.
The monomer dynamics, $g_1(t)$, in the intermediate intra-chain scaling regime also approaches the prediction $g_1(t)\sim t^{\frac{2\nu}{2\nu+1}}$ of the Rouse model (Table~\ref{tab:exponents}). Again, the slow decay of the effective exponents reported in the detailed analysis in Fig.~\ref{fig:RAO_BQT_Mix_scaling}  agrees, at least qualitatively, with the observed crossover of the static exponents in Fig.~\ref{fig:stat_Q_A}.

The situation is more complicated in the melt case.
While the observed monomer MSDs are again in reasonable accord with the prediction $g_1(t)\sim t^{0.4}$ of the Rouse model, the ring CM clearly exhibits anomalous diffusion in the intermediate regime. 
This translates to a scaling of $D \sim N^{-1.6 \pm 0.1}$ over the present range of ring sizes and thus a faster decay than expected from the Rouse model (right column in Fig.~\ref{fig:OverviewDynamics}). We note that
experimental studies on ring polymers in the melt state have also reported a substantial slowdown in COM dynamics ~\cite{Richter:JournalofRheology2021,Kruteva:2023Macromolecules,Schweizer:2023ACSPolym}, which occurs before the system reaches the free diffusion regime.

\subsubsection{Combining monomer and node moves}
\label{res:combo}

The emerging dynamics resulting from the combination of ring monomer and tree node moves in equal proportions ($p_m=0.5 $ in Eq.~\ref{eqn:MCsweep}) are presented in the right column of Fig.~\ref{fig:RAO_BQT_Mix_scaling}.

Given that the node dynamics is significantly faster than the monomer dynamics in ideal and isolated self-avoiding systems (see Fig.~\ref{fig:OverviewDynamics}), it is not surprising that the former dominates the combined dynamics entirely, causing the resulting scaling plots to essentially collapse onto those in the central column.

It is tempting to draw similar conclusions in the melt case, but  with respect to the left column of Fig.~\ref{fig:RAO_BQT_Mix_scaling} and a faster RiAO-like dynamics (bottom row in Fig.~\ref{fig:OverviewDynamics}). Yet, the situation is more subtle since the overal dynamics is accelerated with respect to the pure dynamics of either monomer moves or tree node moves.  
The excellent scaling of the data reported in the bottom right panel of Fig.~\ref{fig:RAO_BQT_Mix_scaling} is thus in itself an interesting observation. In both figures we have tentatively included slopes predicted by the FLG model, but it is difficult to draw any quantiative conclusions at this point.
%
%

\section{DISCUSSION}\label{sec:Discussion}

At this point, we have established a simulation scheme for double-folded ring polymers, which allows us to (i) simultaneously update and trace compatible ring and tree degrees of freedom and (ii) study the emerging dynamics resulting from the combination of different types of local dynamics. 
Our simulations in the context of ideal rings, self-avoiding rings and rings in melts accurately reproduce well-known dynamic properties of three prototypical systems: tightly wrapped rings slithering around fixed trees, rings in an array of obstacles and Brownian motion of trees with quenched connectivity. 

The remaining challenge is to develop a protocol for the analysis of data obtained in mixed dynamical schemes. In particular, this protocol should allow us to identify and distinguish situations where the contributions of different local modes to the emergent dynamics are independent from those where the emergent dynamics arise from coupling effects.


To this end, let us consider two (asymptotically) diffusive processes, $A$ and $B$, in a $d$-dimensional space, generating MSDs $g_A(t)=2d D_A t$ and  $g_B(t) = 2d D_B t$, respectively. 
We are interested in the MSD $g_{AB}(t)$ of mixed schemes, where the local degrees of freedom are updated by $A$ and $B$ moves with probabilities $p_A$ and $p_B$, respectively, and where typically (but not necessarily) $p_A + p_B = 1$. The null model of independent contributions from the $A$ and $B$ dynamics predicts an additive relationship for the resulting diffusion constant:
\begin{equation}
\label{eq_DAB_null_model}
D_{AB}(p_A,p_B) = p_A D_A + p_B D_B
\end{equation}
%
In section~IV in the supplementary material, we propose, in Eq.~(S6), an approximate plausible generalization of Eq.~(\ref{eq_DAB_null_model})
for the full time dependence of the null model, for the monomer and CM MSDs in a mixed scheme. 
In Fig.~6 in section~V in the supplementary material, we show that Eq.~(S6) works well in a case where we expect the null model to apply: the dynamics emerging from the combination of repton and tree node moves, which are indeed completely decoupled within our model.

While the ergodic RiAO dynamics provides an interesting example of the emergence of a qualitatively different dynamics from the combination of local moves (see section~VI in the supplementary material), 
in the remainder of the (main) text, we focus on the emerging dynamics resulting from the combination of monomer and node moves in different proportions, $0\le p_m\le 1$, see Eq.~\ref{eqn:MCsweep}. As evident from Figure~\ref{fig:OverviewDynamics} (in the main text) and Figure~S7 (in the supplementary material), in interacting systems, this combination can lead to an accelerated dynamics compared to the two limiting cases: 
the RiAO dynamics ($p_m=1$) and the dynamics of quenched trees ($p_m=0$).
This raises the question of how this acceleration depends on the ring weight, $N_m$, and the mixing parameter, $p_m$. 

\begin{figure}[H]
\centering
\includegraphics[width=\columnwidth]{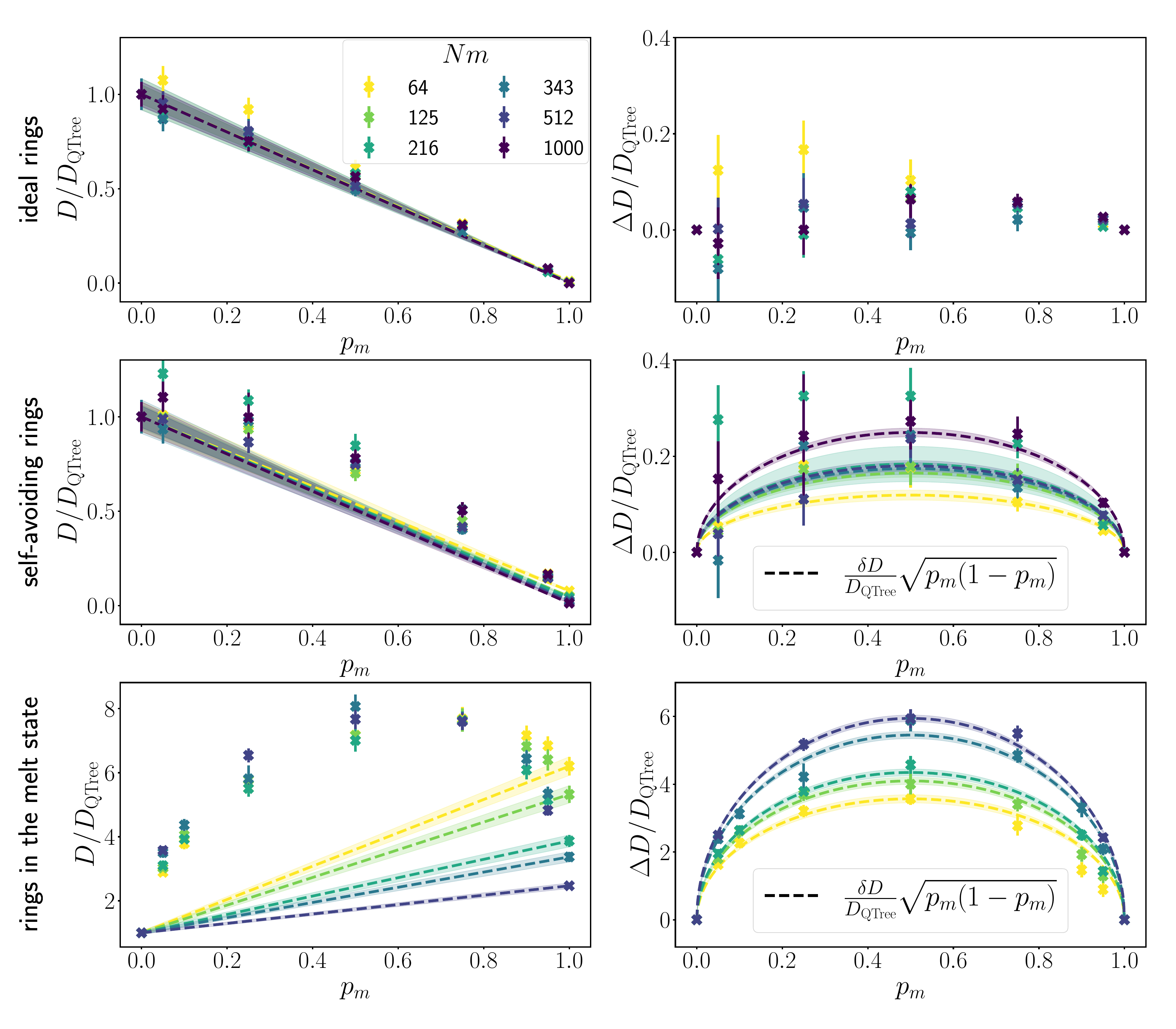}
\caption{Left column: The asymptotic center-of-mass (CM) diffusion constants for different systems, obtained using a combination of monomer and node moves in varying proportions (\(0 \leq p_m \leq 1\)), as described in Eq.~\ref{eqn:MCsweep}. The dashed lines represent the predictions from the null model (Eq.~\ref{eq_DAB_null_model}). Right column: The difference between the observed asymptotic CM diffusion constants and the expected values from the null model, as $\Delta D(p_m) = D(p_m) - \left( p_m D_{\mathrm{RiAO}} + (1-p_m) D_{\mathrm{QTree}}\right)$.  
For better visualization across different system sizes, all data are normalized by \(D_\text{QTree}(N_m)\).}\label{fig:D_Decomposition}
\end{figure}

For our mixing rule, Eq.~(\ref{eqn:MCsweep}), of the monomer and node moves, the prediction of the null model, Eq.~(\ref{eq_DAB_null_model}), reads 
$D(p_m) = p_m D_{RiAO} + (1-p_m) D_{\mathrm{QTree}}$
or, equivalently, 
$D(p_{m})/D_{\mathrm{QTree}} = 1 + (D_{RiAO}/D_{\mathrm{QTree}}-1) p_m$.
The panels in the left column of Fig.~\ref{fig:D_Decomposition} show the corresponding results for the three types of investigated systems (ideal rings, self-avoided rings and rinds in melts). Note that the normalization to $D_{\mathrm{QTree}}$ allows us to show results for rings of different sizes on a common linear scale (Raw results of $D$ are presented in Fig.~S17 and in Table.~II of the Supplementary Material.).

Results reveal the importance and character of deviations of the emergent dynamics from the null model. Namely, while the emergent dynamics in the melt is significantly accelerated and dominated by the deviations from the null model, the corrections appear to be more of a qualitative nature for self-avoiding rings and statistically insignificant for ideal rings. This said, our decision to normalize the diffusion constants in all three cases by those of the {\it asymptotically} faster Brownian Tree Dynamics (right column of Fig.~\ref{fig:OverviewDynamics}) can lead to the wrong impression that $D(p_m)$ extrapolates to zero for finite ring sizes (as it does indeed in the case of the slithering dynamics in Fig.~S6(d) in the supplementary material). However, in the present case $D(p_m)/D_{\mathrm{QTree}}$ remains finite for all $N_m$, albeit much smaller than unity for ideal and self-avoiding double-folded rings and equal to zero only in the limit of infinite chain length.

In the right column of Fig.~\ref{fig:D_Decomposition}, we show the difference\\
\begin{equation}
\Delta D(p_m) = D(p_m) - \left( p_m D_{\mathrm{RiAO}} + (1-p_m) D_{\mathrm{QTree}}\right)
\end{equation}
between the observed asymptotic CM diffusion constant and the expected value from the null model. To simplify a comparison with the data shown in the panels in the left column, we do not show absolute values for $\Delta D(p_m)$ but $\Delta D(p_m)/D_{\mathrm{QTree}}$. If $\Delta D(p_m)$ vanishes by construction in the two limits $p_m=0$ and $p_m=1$ (i.e., a value of zero is in this case a ``true'' and not just an apparent zero), our data suggest an intriguing symmetry in how the node and the monomer dynamics combine.

\begin{figure}[h]
\centering
\includegraphics[width=\columnwidth]{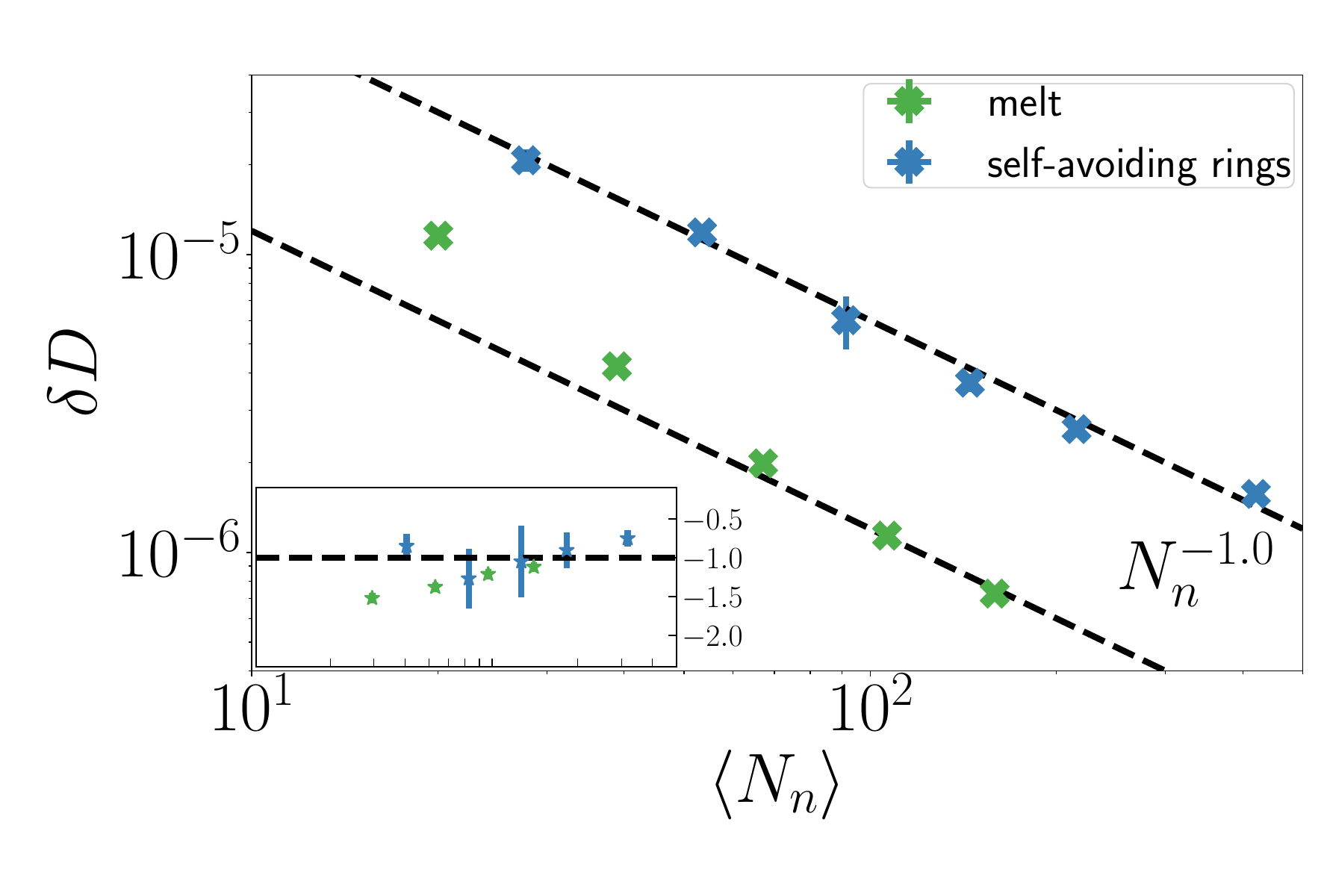}
\caption{Scaling behavior of the coupling contribution $
\delta D$ in Eq.\ref{eq_DAB_total}, extracted from fits of $\Delta D_{AB}(N_n)$ to the functional form $\delta D_{AB}(N_n) \sqrt{p_Ap_B}$. The data suggests a Rouse-like scaling $\delta D (N_n)\sim N_n^{-1}$ for isolated self-avoiding rings, while the behavior for rings in the melt state exhibits a slow crossover towards a similar scaling.}\label{fig:deltaD_scaling}
\end{figure}
Interestingly, while contributions to the ring CM diffusion constant from two coupled A-type and B-type dynamics may take multiple forms (see Sec.~VII in the supplementary material), our results for $\Delta D_{AB}$ reported in the right column of Fig.~\ref{fig:D_Decomposition} suggest that the coupling contribution adopts the simplest possible form of 
\begin{equation}
    \Delta D_{AB} \propto \sqrt{p_A D_A \, p_B D_B} \ .
\end{equation}
Is this (in hindsight) obvious and inevitable or, alternatively, does our observation constrain the choice of admissible microscopic coupling mechanisms?
Since the CM diffusion of our rings is diffusive in the long time limit, we consider as the simplest model the coupling of two diffusion steps, $\vec\delta_A$ and $\vec\delta_B$, for a particle moving under the influence of two random processes $A$ and $B$ such that $\left\langle \left|\vec \delta_A \right|^2 \right\rangle \propto p_A D_A$ and $\left\langle \left|\vec \delta_B \right|^2 \right\rangle \propto p_B D_B$.
To define the additional displacement, $\vec\delta_{AB} = \delta_{AB} \hat\delta_{AB}$, due to the coupling of the two steps, we need to specify its magnitude $\delta_{AB} \equiv \lVert \vec\delta_{AB} \rVert$ and
direction, $\hat\delta_{AB} \equiv \vec\delta_{AB}/\lVert \vec\delta_{AB} \rVert$.
For the magnitude, admissible combinations of $\delta_A$ and $\delta_B$ have to obey the same dimensionality constraint, that we have discussed above for the diffusion constant. With simple vector operations like $\lVert \vec\delta_A \times \vec \delta_B \rVert$ and $\vec\delta_A \cdot \vec \delta_B$ having the unit of the square of a length, one arrives rather naturally at a scaling, $\delta_{AB} \propto\sqrt{\delta_A\delta_B}$,  of the length of the coupling steps with the geometric mean of the elementary step lengths, that may be modulated by geometric factors.
On first sight, this seems to directly lead to the observed scaling $\Delta D_{AB} = \langle \delta_{AB}^2 \rangle \propto \sqrt{\langle \delta_A^2 \rangle\, \langle \delta_B^2 \rangle} \propto \sqrt{p_A D_A p_B D_B}$ of the coupling contribution to the diffusion constant.
However, the relation $\Delta D_{AB} \equiv D_{AB}-p_A D_A-p_B D_B$ only holds, if additionally $\hat\delta_{AB} = \left( \vec\delta_A \times \vec \delta_B \right) / \lVert \vec\delta_A \times \vec \delta_B \rVert$ or if the direction of the coupling steps, $\hat\delta_{AB}$, is not correlated with $\vec\delta_A$ and $\vec\delta_B$. In contrast, if $\hat\delta_{AB}$ lies in the plane spanned by the elementary steps, $\vec \delta_A$ and $\vec \delta_B$, correlations lead to more complicated relations of the general type Eq.~(S13) in the supplementary material.

As our final point, 
we address the extrapolation of our data to ring sizes beyond the range studied in the present work. 
 On first sight, the diffusion constants reported in Fig.\ref{fig:OverviewDynamics} seem compatible with three power laws for the two pure and the mixed dynamics with $p_m=0.5$. In particular, the data for the total diffusion constant in the mixed case are close to the prediction of the FLG model~\cite{Rubinstein:Macromolecules2016}.
However, this interpretation is incompatible with our observations for the full range of $p_m$ values shown in Fig.\ref{fig:D_Decomposition}. In particular, for other values of $p_m$ our data for $D$ do not follow a uniform power law (see Fig.~S16 in the Supplementary Material). Instead, our analysis suggests that the total diffusion constant for the mixed dynamics can be expressed as:
\begin{equation}
\label{eq_DAB_total}
D_{AB} = p_A D_A + p_B D_B + \delta D_{AB} \sqrt{p_A p_B}
\end{equation}
where
\begin{equation}\label{eq:delta_AB}
    \delta D_{AB}(N) \equiv \frac{\Delta D_{AB}(N)}{\sqrt{p_A p_B}}= a(N)\sqrt{D_A(N) D_B(N)} 
\end{equation}
is the magnitude of the contribution to the diffusion constant arising from the coupling.

In spite of the vastly different values of $D_{\mathrm{QTree}}$ for isolated self-avoiding rings and for rings in a melt being approximately two orders of magnitude smaller (As shown in Fig.\ref{fig:OverviewDynamics} (main text) and in Table~II (Supplementary Material)), Fig.~\ref{fig:deltaD_scaling} shows that the absolute magnitude of the $\delta D$ arising from the coupling with the RiAO-like dynamics is quite similar for the two systems. Moreover, in both cases $\delta D$ appears to be of the order of $ D_{\mathrm{QTree}}$, i.e. the dimensionless prefactor $a(N)$ in Eq.~(\ref{eq:delta_AB}) for the ring-size dependent efficiency of the coupling is by no means a trivial constant.
For isolated self-avoiding rings $\lim_{N\rightarrow\infty} \delta D /D_{\mathrm{QTree}} > 0.4$ and
our data are in good agreement with a Rouse-like scaling, $\delta D\sim N^{-1}$.
Our data for melts are less clear. While they show no sign of converging to the FLG prediction of $\delta D\sim N^{-5/3}$, they are compatible with a slow crossover to a Rouse-like scaling.
In particular, $\lim_{N\rightarrow\infty}\delta D /D_{\mathrm{QTree}} > 12$.
Notably the total CM diffusion constant in the melt case seems to be dominated asymptotically by the mobility arising from the coupling of the Brownian tree and the RiAO-like dynamics.

%
%
\section {SUMMARY AND CONCLUSION }\label{sec:CONCLUSION}
Ring polymers pose a number of challenging fundamental problems in polymer physics~\cite{Kruteva:2023Macromolecules,Schweizer:2023ACSPolym,Mei_PNAS:2024}.
In this article, we have studied the dynamics of a (lattice) model of tightly double-folded chains \cite{Khokhlov1985Phy.L.A, Rubinstein1986PhysRevLett., Obukhov.Rubinstein1994PRL,SmerkGrosberg2015Cond.Matt,ElhamPRE2021}.
In our Monte Carlo simulations, ring monomer moves are responsible for the spontaneous creation and deletion of side branches and the diffusive local mass transport along the tree, while the exclusive application of tree node moves gives rise to a Brownian dynamics of double-folded rings with quenched secondary structure.
In particular, the present generalization of our earlier model \cite{ElhamPRE2021} allows us to explore arbitrary combinations of ring and tree dynamics where sets of compatible ring and tree degrees of freedom are simultaneously updated and defined at all moments of time. 

As a starting point, we have explored the dynamics emerging from different combinations of local moves for three paradigmatic types of tree-like double-folded ring systems: ideal rings, isolated self-avoiding rings, and rings in the melt state. 
The local transport of reptons gives rise to a slithering of rings around a fixed tree, while the ring-in-an-array-of-obstacles (RiAO) dynamics emerges in the presence of additional monomer moves, which allow to absorb and grow side branches. 
The QTree dynamics of quenched trees undergoing Brownian motion turns out to be Rouse-like for ideal and isolated self-avoiding rings. In the melt state, the monomer diffusion is again in good agreement with the expectation $g_1\sim t^{0.4}$ from the Rouse model for crumpled rings. However, there are also clear indications for a slowdown of the center-of-mass dynamics on intermediate scales, which translates into a reduced center-of-mass diffusion $D_{\mathrm{QTree}}\sim N^{-1.6\pm0.1}$.

To apply the model to a particular ring polymer system (see~e.g.~\cite{Vologodskii_review1994,Junier:2023FrontMicrobiol,RNALiu2016,vRNAKellyGrosberg2016,PrueferRNASingaram2016,Rosa2008PLOS,LiebermanAiden2009.Science,Halverson:Rep.Prog.Phys2014}), 
one needs to specify the relative importance of the different types of local dynamics represented by the ring monomer and tree node moves.
For this purpose we have introduced a control parameter ($0\le p_m\le 1$) allowing us to explore the full range of dynamical scenarios, where the limiting cases are RiAO  ($p_m = 1$) and QTree ($p_m = 0$) dynamics. The appropriate value of  $p_m$ for a physical system reflects the microscopic mobility of individual monomers relative to the collective mobility of tree-like segments. These mobilities depend on several physical factors, including local friction, chain stiffness, and topological constraints.
For example, in RNA molecules, the secondary structure typically corresponds to a quenched tree-like system. In contrast, dense polymer melts may be better described by RiAO dynamics. For small, supercoiled DNA molecules, an intermediate value of $p_m$ is more likely, as both slithering dynamics~\cite{Huang_Dynamics_2001} and the creation/annihilation of branches~\cite{chirico_brownian_1996,van_loenhout_dynamics_2012} contribute to the overall behavior. By comparing MSDs measured in physical or biological systems, one should be able to infer the value of $p_m$ that best represents the dynamics at play.

Our objective here was to study generic aspects of the coupling of different local modes of dynamics. In particular, we have exploited the opportunity inherent in a mesoscale description to freely vary the relative kinetic rates instead of being limited to ratios emerging in particular experimentally or numerically accessible target or model systems. 
For non-interacting ideal rings we found that the dynamics emerging from the simultaneous presence of monomer and node moves are well described by an additive null model. In contrast, a non-trivial acceleration emerges for interacting rings, which we have analyzed in detail in Sec.~\ref{sec:Discussion}. 
For self-avoiding trees, the corresponding contribution to the ring center-of-mass diffusion constant scales with the ring size like $\delta D\sim N^{-1}$. This corresponds to a sub-dominant acceleration of the Rouse-like QTree dynamics.
In the melt case, the coupling contribution dominates. While the scaling is less clear, our data are compatible with the interpretation that constraint release due to the annealing of the tree structure asymptotically induces a Rouse-like dynamics.


Finally, we note that numerous experimental investigations using fluorescence microscopy have explored the 3D dynamics of chromatin in the cell nuclei of mammals~\cite{Zidovska:2013Nat.AcademySciences}, \textit{Drosophila}~\cite{MARSHALL:1997CurrentBiology,VAZQUEZ:2001CurrentBiology}, and yeast~\cite{MARSHALL:1997CurrentBiology,Cabal:2006Nature} as well as in the bacterial nucleoid~\cite{Espeli:2008DNA,Weber:2010PhysRevLett.,Weber:2012NationalAcademyofSciences,Javer:2013Nat.Comm.}. Reported results show a variety of subdiffusive behaviors, with $g_1$ exponents typically ranging from $\sim 0.3$ to $\sim 0.7$, depending on cell activity, environment, and the specific chromatin locus under study.
In this context, our observation of an exponent $\simeq 0.4$ in the dynamics of monomers in dense ring melts is strikingly reminiscent of the typical exponent of $0.4$ observed in bacterial chromosomes~\cite{Weber:2010PhysRevLett.,Weber:2012NationalAcademyofSciences,Javer:2013Nat.Comm.}. While several scenarios have been proposed to explain this value (see, e.g.~\cite{Subramanian:2023Phys.Rev.Res.} and references therein), it is notable that a tree-like, double-folded ring corresponds to the typical structure associated with supercoiled DNA, a hallmark of bacterial chromosomes~\cite{Vologodskii_review1994,Junier:2023FrontMicrobiol}. Nonetheless, this correspondence should be interpreted with care. Our study concerns Brownian-like, near-equilibrium situations where the only source of energy is the thermal activity of the (implicit) surrounding medium. In contrast, chromatin dynamics \textit{in vivo} is shaped by many active, far-from-equilibrium processes such as ATP-dependent remodeling, transcriptional activity, DNA replication, and loop extrusion~\cite{MARKOpre2015,Junier:2023FrontMicrobiol,Harju:2025Physical}. Additionally, chromatin exhibits structural and topological complexity that is not captured by our models, and the relevance of our results may depend on the time and length scales considered. In this regard, observations in eukaryotes 
\cite{BrucknerScience:2023,MachNatureGenetics:2022} of exponents in the range of $0.5 - 0.6$ beyond the entanglement or branching scale call for different explanations than those advocated in the present work.

%
%

\section{SIMULATION SOFTWARE}\label{sec:software}

The simulation software is an open-source package, distributed under the GNU General Public License v$3.0$ and written in $\textnormal{C++}$. It is available at \cite{ghobadpour_2025_gitlab_page}. All simulations were performed on CBPSmn computer cluster of ENS-Lyon. 

%
%

\section{ACKNOWLEDGMENT}
We gratefully acknowledge the support of the Centre Blaise Pascal's IT test platform at ENS de Lyon (Lyon, France) for facilities. The platform operates the SIDUS solution \cite{Quemener:CBP2013} developed by Emmanuel Quemener. 
We thank Angelo Rosa and Cedric Vaillant for valuable discussions, and Ali Farnudi for his technical support. IJ and RE acknowledge funding from CNRS 80 Prime (MIMIC project). RE has greatly benefited from discussions with Michael Rubinstein, Dimitris Vlassopoulos and Charles M. Schroeder at and after the 2023 Cecam workshop on ring dynamics in Prato.

%
%

\section{SUPPORTING INFORMATION}\label{SM}

Supporting Information includes:
\begin{itemize}

\item \textbf{Sec.~I:} Equilibration
\item \textbf{Sec.~II:} Effect of maximal functionality of tree nodes on the dynamics
\item \textbf{Sec.~III:} Stored-Length (Repton) Distribution
\item \textbf{Sec.~IV:} Approximate solution of the null model for the dynamics resulting from independent additive contributions
\item \textbf{Sec.~V:} Comparison of the null model to simulation data for double-folded rings 
\item \textbf{Sec.~VI:} The emergence of the ring-in-an-array-of-obstacles-like (RiAO) dynamics from the combination of hairpin and repton moves
\item \textbf{Sec.~VII:} Possible contributions to the ring CM diffusion constant from two coupled dynamics
\item \textbf{Sec.~VIII:} Brownian dynamics of randomly quenched trees (QTree dynamics)
\item \textbf{Sec.~IX:} Ring-in-an-array-of-obstacles-like  (RiAO)dynamics

\item \textbf{Table I:} Summary of system parameters for polymer chains.
\item \textbf{Table II:} Diffusion constants $D$ for different systems, sizes, and mixing ratios.

\item \textbf{Figure S1:} Equilibration analysis.
\item \textbf{Figures S2-S3:} Effect of maximum node functionality on structural and dynamical properties.
\item \textbf{Figures S4:} Distribution of tree node functionalities for systems with different imposed maximal functionalities ($f_{max} = 3,6,9,12$). 
\item \textbf{Figure S5:} Stored-length (repton) distribution and Poisson model.
\item \textbf{Figure S6:} MSDs for ideal double-folded rings, resulting from the combination of slithering dynamics and quenched tree dynamics, as a function of the mixing fraction $p_m$.
\item \textbf{Figure S7:} MSDs resulting from the superposition of Quenhced Tree (QTree) dynamics and the Ring-in-an-array-of-obstacles-like (RiAO) dynamics, based on the mixing fraction $p_m$.
\item \textbf{Figure S8:} The MSDs and the difussion constants for varying the relative weights of repton and hairpin move in the ring-in-an-array-of-obstacles-like (RiAO) dynamics.

\item \textbf{Figures S9--S11:} Brownian quenched tree dynamics (Ideal, Self-avoiding, Melt systems).
\item \textbf{Figures S12--S14:} Ring-in-an-array-of-obstacles-like (RiAO) dynamics (Ideal, Self-avoiding, Melt systems).
\item \textbf{Figure S15:} Slithering dynamics of ideal rings.
\item \textbf{Figure S16:} Scaling of the CM diffusion constant $D$ for varying mixing ratios $p_m$.
\item \textbf{Figure S17:} Relaxation times and CM diffusion constants versus $p_m$ compared to null model.

\end{itemize}

\renewcommand{\thesection}{}
\renewcommand{\thesubsection}{}
\addcontentsline{toc}{section}{REFERENCES}
\bibliography{bibliography}

\end{document}